# BEAM-BEAM ISSUES FOR COLLIDING SCHEMES WITH LARGE PIWINSKI ANGLE AND CRABBED WAIST

Pantaleo Raimondi[1], Dmitry Shatilov[2], Mikhail Zobov[1]

*[1)]INFN-Laboratori Nazionali di Frascati Via E. Fermi 40, I-00044 Frascati, Italy*
*[2)]BINP, Lavrentieva av. 11, Novosibirsk 630090, Russia*

## Abstract

Numerical simulations have shown that a recently proposed “crabbed waist” scheme of beam-beam collisions can substantially increase the luminosity of a collider. In this paper we give a qualitative explanation why this scheme works. For this purpose we use simple geometrical considerations and analyze peculiar properties of vertical motion modulations by synchrotron and horizontal betatron oscillations. It is shown that in the "crabbed waist" scheme these modulations, which are the main sources of beam-beam resonances excitation, are significantly suppressed. Some numerical examples demonstrating the effect of the crabbed waist collisions are also given.

## 1 LUMINOSITY CONSIDERATIONS

The recently proposed Crabbed Waist (CW) scheme[1)] of beam-beam collisions can substantially increase collider luminosity since it combines several potentially advantageous ideas. The first one is large Piwinski's angle. For collisions under a horizontal crossing angle $\theta$ (flat beams) the luminosity $L$, the horizontal $\xi_x$ and the vertical $\xi_y$ tune shifts scale as[2)]:

$$L \propto \frac{N \cdot \xi_y}{\beta_y}; \qquad \xi_y \propto \frac{N \cdot \beta_y}{\sigma_x \sigma_y \cdot \sqrt{1+\phi^2}}; \qquad \xi_x \propto \frac{N}{\varepsilon_x \cdot (1+\phi^2)} \tag{1}$$

Here Piwinski's angle $\phi$ (see Fig.1) is defined as:

$$\phi = \frac{\sigma_z}{\sigma_x} tg\,\theta \approx \frac{\sigma_z}{\sigma_x}\theta \tag{2}$$

where $\sigma_{x,y,z}$ are the r.m.s. bunch sizes, $\beta_y$ is the vertical beta-function at the IP, $\varepsilon_x$ is the horizontal betatron emittance, $N$ is the number of particles per bunch.

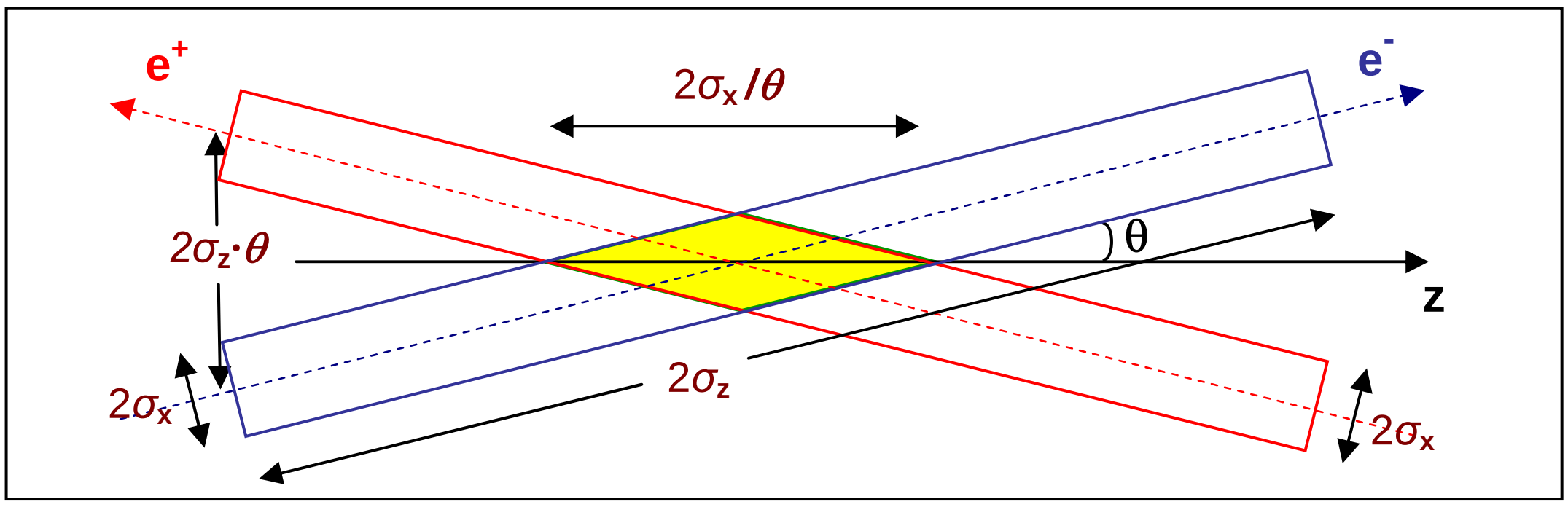

**FIG. 1:** Scheme of collision with a crossing angle.

As it is seen from (2), there are four possibilities of increasing Piwinski's angle:

1) Increasing the bunch length $\sigma_z$. According to (1-2) this allows to increase $N$ proportionally to $(1+\phi^2)^{1/2}$. In this case $\xi_y$ remains constant, $\xi_x$ decreases and $L$ increases proportionally to $N$. The overlap area of colliding bunches does not change.

2) Increasing the crossing angle $\theta$ (but in assumption it remains to be << 1). The consequences are actually the same as in the previous case, but the overlap area of colliding bunches decreases as $1/\theta$.

3) Decreasing the horizontal emittance $\varepsilon_x$ and, therefore, the horizontal beam size $\sigma_x$. In this case, if $\phi$>>1, we should keep the same $N$, so that all the values $\xi_x$, $\xi_y$, $L$ do not change. The only effect is that the overlap area of colliding bunches decreases. However, if we go from small $\phi \leq 1$ to large values, we need to decrease $N$ in order to keep the $\xi$ values, so the luminosity even drops slightly. On the other hand, if one (or both) of the 1–2 possibilities were also exploited ($\xi_x$ decreased), we can afford some $\xi_x$ increase due to $\varepsilon_x$ change, while $\xi_y$ can be controlled by the betatron coupling, so we can keep N and L unchanged.

4) Decreasing the horizontal beta-function $\beta_x$ and, therefore, the horizontal beam size $\sigma_x$. In this case, if N is increased proportionally to $(1+\phi^2)$, $\xi_x$ will not change, but $\xi_y$ will grow proportionally to N (assuming that $\sigma_x \cdot (1+\phi^2)^{1/2} = const$, for $\phi >> 1$). Or, if we do not change N, it simply results in $\xi_x$ decrease.

As usual, $\xi_y$ is more "flexible" parameter as it can be controlled by the coupling (change of $\sigma_y$), so the main limit on the beam current N is set by the horizontal tune shift $\xi_x$. On the other hand, using the 1-2) and 4) features we can decrease $\xi_x$ significantly, far below the beam-beam limit, while the coupling cannot be changed too much: very small values are difficult to achieve, big values result in the vertical aperture (in units of $\sigma_y$) decrease. Thus, we can conclude that the 1) and 2) features easily allow the N and L increase proportionally to $(1+\phi^2)^{1/2}$, while 3) and 4) can allow some N and L increase, but it is more questionable and requires the coupling increase, shrinking the vertical aperture (in units of $\sigma_y$). So, the main advantages of the 3) and 4) features are the overlapping area decrease, plus $\xi_x$ decrease by 4). In the proposed CW scheme[3,4)] the crossing angle is increased, but not significantly – by a factor of 2-3 as compared to the currently working meson factories (DAFNE and KEKB). The main accent is made on the significant $\sigma_x$ decrease. Thus we do not get very essential luminosity gain when only increasing the Piwinski's angle, but it opens the possibility for the next step (second idea): significant $\beta_y$ decrease.

It is well known that decreasing $\beta_y$ at the IP is very profitable for the luminosity (see Eq.1). In assumption that we should have the same beam-beam tune shift $\xi_y$, the $\beta_y$ decrease allows the N increase as $1/\beta_y^{1/2}$ (if $\xi_x$ allows, that is just the case for large Piwinski's angles), so the luminosity goes up as $1/\beta_y^{3/2}$. But in order to get all these advantages we (usually) need to keep the bunch length not larger than $\beta_y$. So, in ordinary IR concepts the main limitation on further $\beta_y$ decrease is set by the lower limit on the achievable bunch length $\sigma_z$. But more precisely, the condition is that $\beta_y$ should be comparable to the *overlapping area*, not the bunch length! Usually (head-on or small $\phi$) these two are comparable, but with large Piwinski's angle the overlapping area becomes much smaller than $\sigma_z$, allowing significant $\beta_y$ decrease:

$$\beta_y \approx \frac{\sigma_x}{\theta} << \sigma_z \qquad (3)$$

And this can give us very significant gain in the luminosity! The additional advantages of such collision scheme are:

- There is no need to decrease the bunch length to increase the luminosity as proposed in standard upgrade plans for B- and Φ- factories[5,6,7)]. This will certainly ease the problems of HOM heating, coherent synchrotron radiation of short bunches, excessive power consumption, etc.

- The problem of parasitic collisions (PC) is automatically solved since with higher crossing angle and smaller horizontal beam size the beams separation at the PC is very large in terms of $\sigma_x$.

On the other hand, such a scheme of collision induces strong X-Y betatron resonances, which may cause troubles in making choice of the working point, and lower the achievable luminosity. Fortunately, a very attractive and simple solution was proposed recently[1)], which solves this problem – the Crabbed Waist (CW). One of the reasons it was proposed for is the geometrical luminosity increase, but calculations[8)] showed that for the typical beam parameters the gain in luminosity is relatively small, of the order of 5-10%. So, the main profit of the CW comes from the beam dynamics considerations – the betatron resonances suppression. This is the third idea accomplishing the new concept of the IR design.

## 2 BEAM DYNAMICS CONSIDERATIONS

Here we qualitatively consider the main features of different collision schemes concerning beam-beam interaction, in particular how the vertical particle's motion is affected by the synchrotron and horizontal betatron oscillations. The "luminosity" aspect is not considered here. What is important in the "old concept" ($\phi \leq 1$): *where the particle meets the center of the opposite bunch*. This is what we call CP (Collision Point), while the IP (Interaction Point) is the nominal CP for an equilibrium particle. The main features of the old concept are:

a) The X-Y betatron resonances appear due to the vertical beam-beam kick's dependence on the horizontal particle's coordinate (amplitude modulation), see Fig.2. The horizontal kick also depends on the vertical coordinate, but for the flat beams this dependence is much weaker.

b) Without crossing angle the SB resonances appear by two reasons. The first one is the betatron phase advance modulation from IP to CP, especially for the vertical betatron motion (assuming $\beta_y$ is small, flat beams). The second one is the $\xi_y$ increase at the CP (that is an amplitude modulation), when shifted longitudinally from IP, due to the hour-glass effect, see Fig.3 (red line).

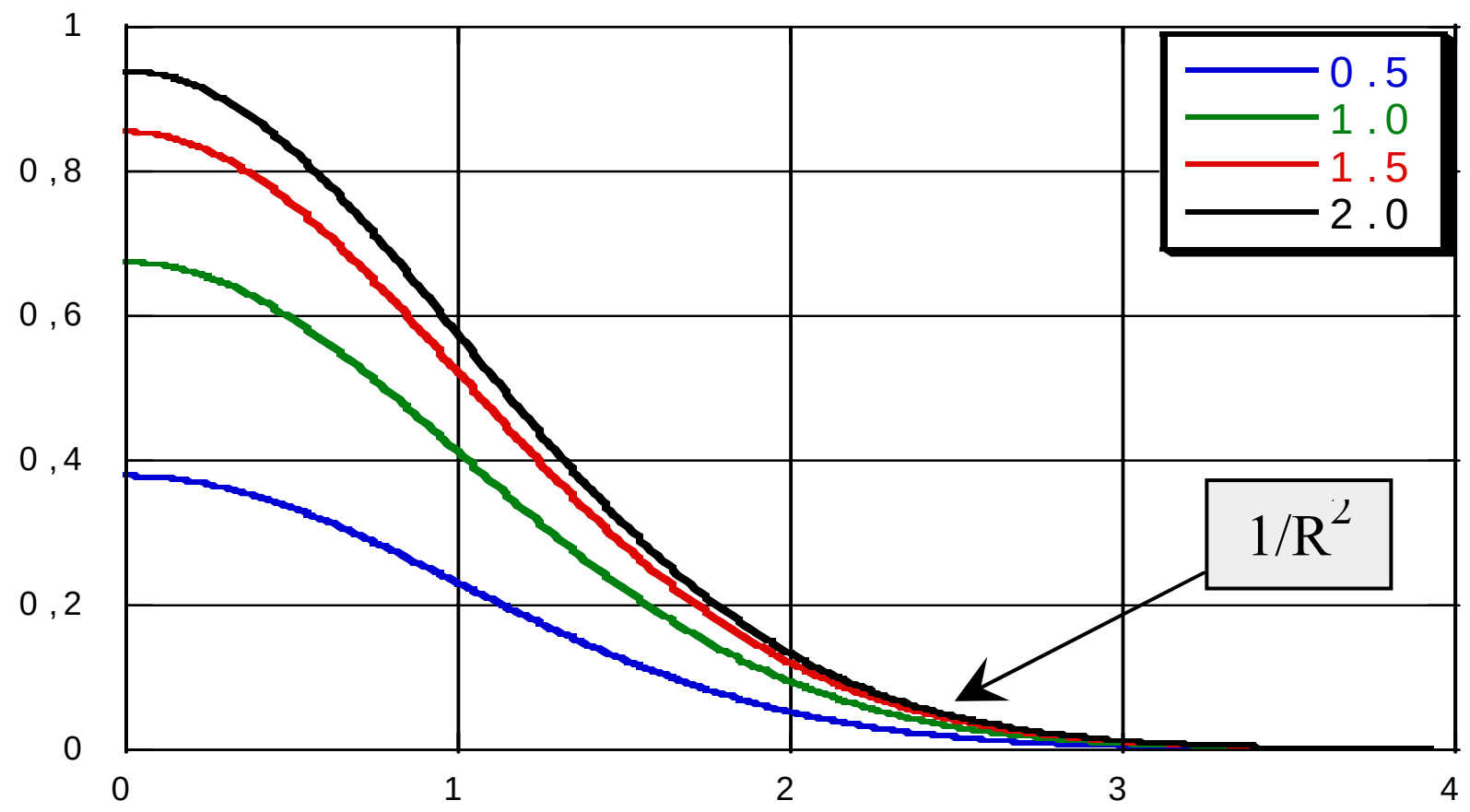

**FIG. 2:** Vertical kick vs. $X/\sigma_x$ for 4 different $Y/\sigma_y$: 0.5, 1.0, 1.5, 2.0. Bassetti-Erskine formula, $\sigma_x/\sigma_y = 100$.

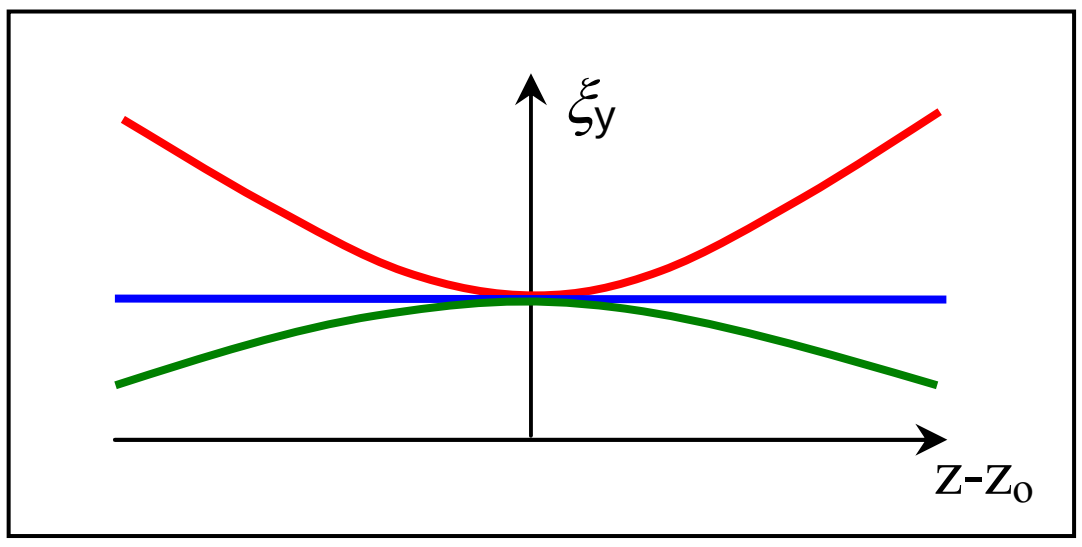

**FIG. 3:** Synchrotron modulation of $\xi_y$, qualitative picture. Two lines correspond to head-on collisions: flat beams (red) and round beams (blue). Green line corresponds to the crossing angle collisions, flat beams.

With crossing angle (horizontal, small Piwinski's angle, see Fig. 4) we have:

c) Horizontal coordinate of the test particle in CP (in the strong bunch's coordinate frame) depends mainly on the particle's longitudinal coordinate, that results in a strong amplitude modulation of both horizontal and vertical beam-beam kicks by the synchrotron oscillations. This can excite strong synchro-betatron resonances, both horizontal and vertical.

d) Particles with non-zero X-coordinate have the CP shifted longitudinally. This is similar with the effect of synchrotron oscillations, but induces X-Y betatron resonances instead of synchro-betatron ones. However, this effect is rather small for $\phi \leq 1$, since the longitudinal shift of CP due to X-oscillations is not significant, much smaller than $\beta_y$.

e) $\xi_y$ decreases at the CP (when shifted longitudinally from IP) due to horizontal separation, see Fig.3. Although the $\xi_y$ decreasing could seem better than increasing, it is in any case an amplitude modulation and results in synchro-betatron resonances excitation.

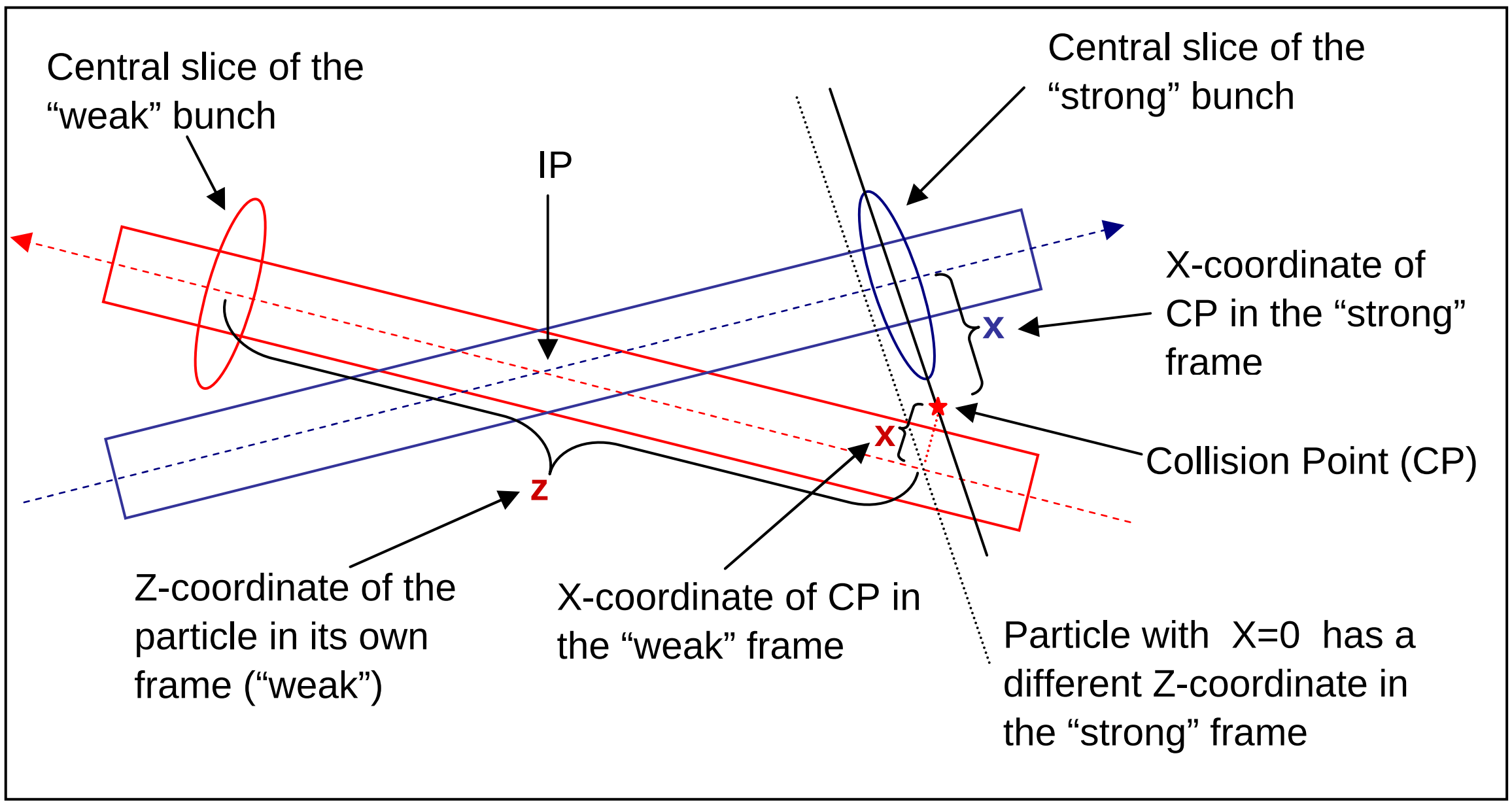

**FIG. 4:** Collision with a crossing angle, $\phi \leq 1$.

In the "new concept" (large Piwinski's angle) the situation is rather different. The center of the opposite bunch is not so important and can be not seen at all (due to large horizontal separation) by the particles with large longitudinal displacements. So, the CP has to be defined in a different way: now it is the point where the test particle crosses the longitudinal axis (i.e. trajectory of the equilibrium particle) of the opposite beam, see Fig. 5. The main features of the new concept (step by step) are:

### 2.1 Large Piwinski's angle (small overlapping area, but still $\beta_y \sim \sigma_z$)

1) Particle interacts with a small part of the opposite bunch, near the IP. When considering the vertical beam-beam kick, the rest of the bunch is not seen due to large horizontal separation. The horizontal kick, however, is more "long-distance". Indeed, for large horizontal separations the horizontal kick strength drops as 1/r, while the vertical one – as $1/r^2$. But in general, as our main concern is the vertical motion, and since the overlapping area (OA) is much smaller than $\beta_y$, we can consider CP coincides with the nominal IP. Since the particle's betatron phase and the opposite beam's parameters at the CP do not depend on the particle's longitudinal and horizontal coordinates, the effects of (b) and (d) vanish.

2) The vertical beam-beam kick's dependence on the particle's X-coordinate becomes much smaller than in the ordinary IR design, – almost negligible, since the particle shifted horizontally crosses the opposite bunch in the point slightly shifted longitudinally, but with actually the same density, and the geometry of collision will be the same as for the equilibrium particle, thus eliminating the (a) effect. This makes the X-Y betatron resonances much weaker than even in the ordinary head-on collisions! So, the beam-beam interaction can be considered, in some sense, as one-dimensional.

3) The synchrotron motion still affects the beam-beam interaction through the modulation of the density of the opposite bunch at the CP: the particle shifted longitudinally will meet at the IP the other part of the opposite bunch, also shifted longitudinally, and therefore having the different density. This is the only modulation remaining.

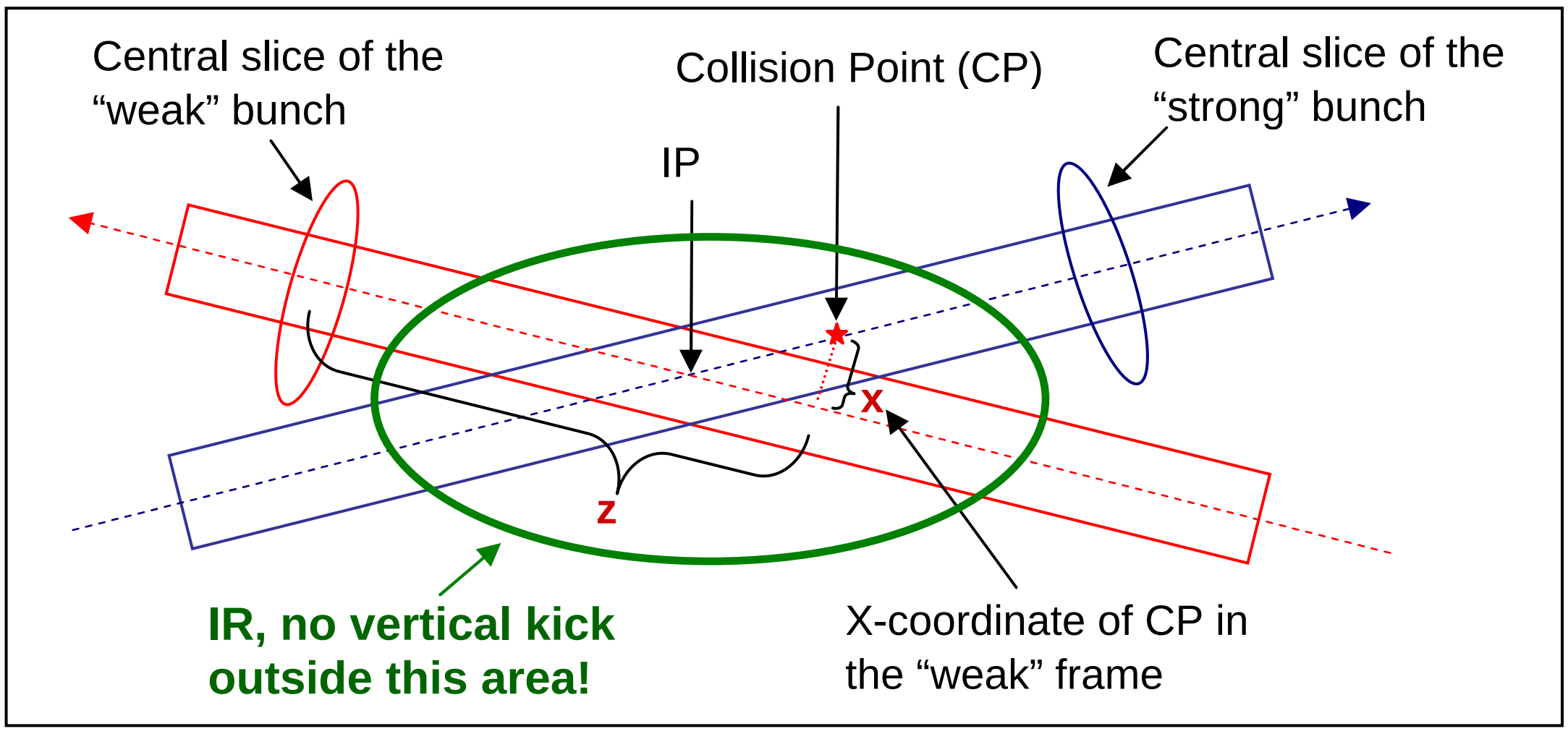

**FIG. 5:** Collision with a crossing angle, $\phi >> 1$.

4) On the other hand, the betatron phase advance over the OA is small, so a particle fills a "solid" kick in a constant phase. This makes the beam-beam interaction even stronger as compared to the normal situation, where the betatron phase averaging during the interaction makes the beam-beam kick "smoother".

### 2.2 Small $\beta_y$ at the IP (to fit the OA)

5) The vertical betatron phase advance from IP to CP becomes significant again. So, the betatron phase modulation appears again and becomes much stronger than in (d) case, since the longitudinal shift of CP due to horizontal betatron oscillations is comparable now with $\beta_y$, see Fig. 6.

6) Amplitude modulation of the vertical beam-beam kick (i.e. $\xi_y$) by the horizontal betatron oscillations also appears, and has two sources: $\beta_y$ of the "weak" beam in the numerator and $\sigma_y$ (that is proportional to $\beta_y^{1/2}$) of the "strong" beam in denominator.

7) The phase averaging over the interaction region appears again, smoothing the beam-beam kick. This leads also to a strong suppression of the vertical synchro-betatron resonances[9].

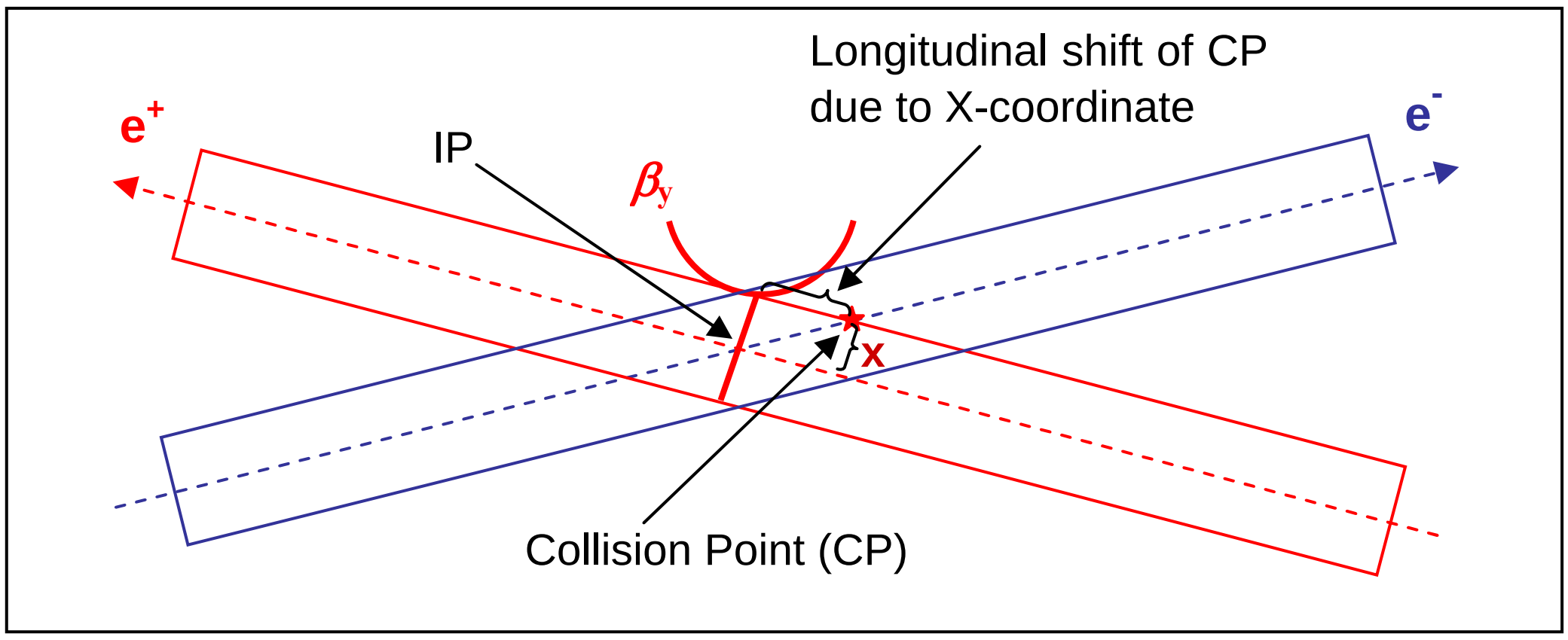

**FIG. 6:** Small $\beta_y$, collision with a crossing angle, $\phi$ >> 1.

### 2.3 Crabed waist (see Fig.7)

8) When applied to the "weak" beam, it removes the betatron phase modulation at the CP (see explanations in the next chapter). Besides, the amplitude modulation also changes: $\beta_y$ of the "weak" beam in CP ($\xi_y$ numerator) is not modulated anymore, but modulation of $\sigma_y$ ("strong" beam, $\xi_y$ denominator) still remains.

9) When applied to the "strong" beam, it somehow affects the beam-beam kick's dependence on the particle coordinates, so that Bassetti-Erskine formulae cannot be used more. It is not clear yet how it could affect the beam-beam resonances, whether it is profitable or not, but we believe it is much less significant than suppression of the betatron phase modulation.

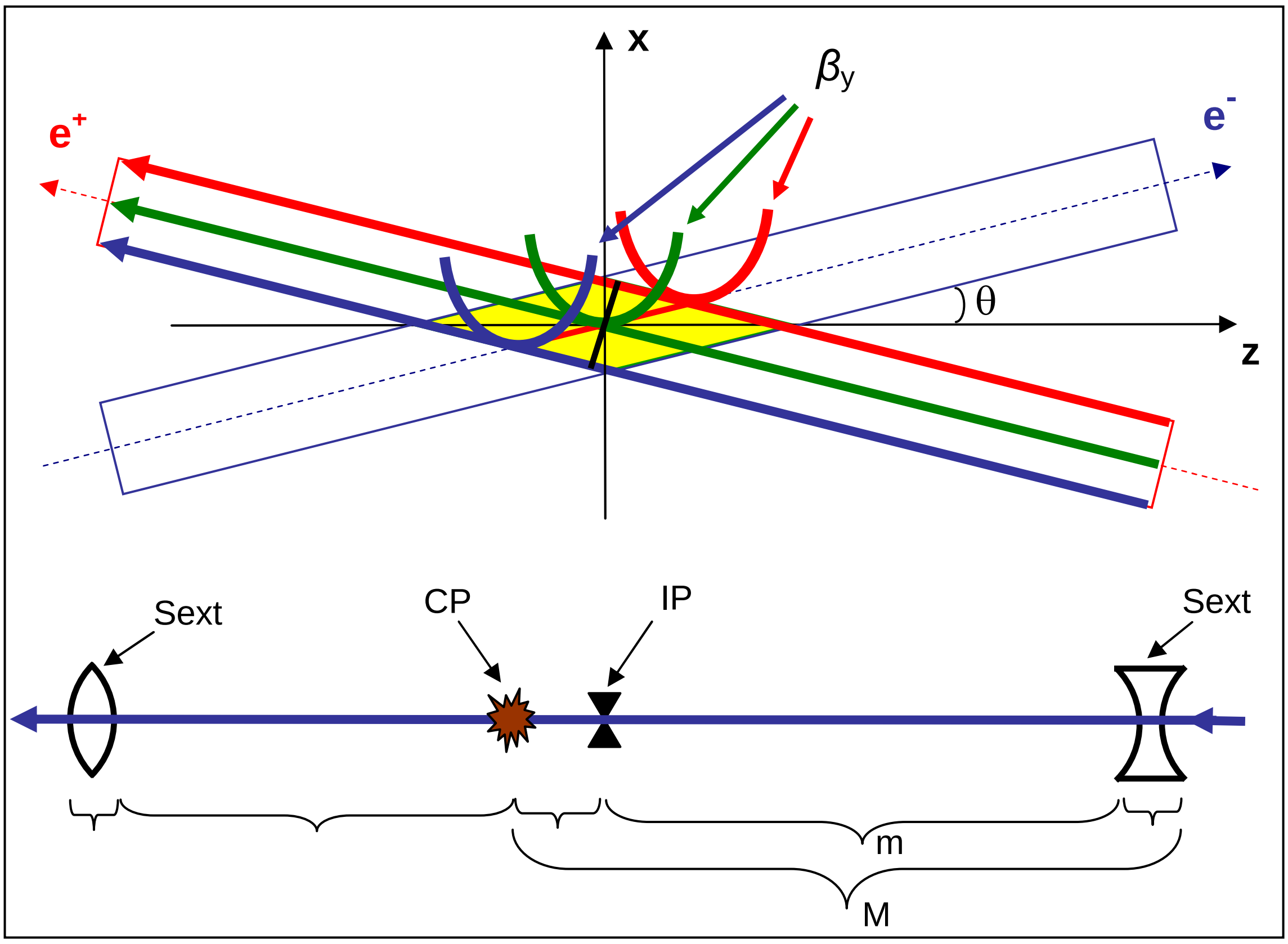

**FIG. 7:** Collision in a “Crabbed Waist” scheme.

## 3 BETATRON RESONANCES SUPPRESSION BY THE CRABBED WAIST

Let us consider in more details how the Crabbed Waist (CW) works. According to the definition, it means that for the particles shifted horizontally (betatron X-coordinate is not zero) the minimum of the vertical beta-function (i.e. the vertical waist) is shifted longitudinally. The necessary nonlinear transformation can be provided by two sextupoles placed on both sides of the IP. The horizontal and vertical phase advances between the sextupoles and the IP must satisfy the following conditions:

$$\begin{aligned} \Delta\mu_x &= \pi \cdot m \\ \Delta\mu_y &= \pi / 2 \cdot (2n+1) \end{aligned} \tag{4}$$

where *m, n* are integers. The required sextupole strength is:

$$\mathrm{K} = \frac{\sqrt{\beta_x^* / \beta_x}}{2\theta \cdot \beta_y^* \cdot \beta_y} \tag{5}$$

where “*” denotes the beta-functions at the IP, without “*” – at the sextupole location. The two sextupoles must have the different polarity. If the (4) conditions are satisfied, they exactly compensate each other, so all the perturbations induced by them become local. In these conditions we can say that *between the sextupoles* $\beta_y$ *is a function of the horizontal betatron coordinate X*. So, for a particle with any given X-coordinate we can consider the sextupoles as additional linear lenses, as if they are a part of linear lattice. What is very important: in such a lattice the condition (4) for $\Delta\mu_y$ is valid for the phase advance between the sextupoles and the CP (not IP)! Indeed, the transport matrix M (see Fig. 7) from the entrance of the first sextupole to the CP (vertical betatron motion only) can be written as:

$$M = \begin{pmatrix} 1 & L \\ 0 & 1 \end{pmatrix} \cdot \begin{pmatrix} m_{11} & m_{12} \\ m_{21} & m_{22} \end{pmatrix} \cdot \begin{pmatrix} 1 & 0 \\ V & 1 \end{pmatrix} \tag{6}$$

where the first matrix corresponds to the drift space from IP to CP, L being the drift length, the last matrix corresponds to the sextupole, considered here as a thin linear lens, and in the middle is the unperturbed matrix m from the sextupole location to the IP. It is important to note that $m_{22} = 0$, since $\alpha_y = 0$ at the IP and $\Delta\mu_y = \pi/2$. As a result, for M we get this matrix element equals to zero too: $M_{22} = 0$. On the other hand, considering the "new" lattice (sextupoles included) we can write the standard formula for $M_{22}$:

$$M_{22} = \sqrt{\beta_y / \beta_{1y}} \cdot \left(\cos(\Delta\mu_{1y}) - \alpha_{1y} \cdot \sin(\Delta\mu_{1y})\right) \tag{7}$$

where $\beta_{1y}$ and $\alpha_{1y}$ are the beta- and alpha-functions at the CP. Since it is the waist at the CP, $\alpha_{1y}$ must be equal to zero, so we get $\cos(\Delta\mu_{1y}) = 0$, resulting in $\Delta\mu_{1y} = \pi/2$, that is exactly what we wanted. In the other words, the vertical betatron phase advance from the first sextupole to CP and then from CP to the second sextupole remains to be $\pi/2$ for all the particles independently on their X-coordinate, thus eliminating the vertical betatron phase modulation by the horizontal betatron oscillations.

Now let us consider amplitude modulation of the vertical beam-beam kick caused by the $\beta_y$ modulation at the CP. The vertical tune shift depends on both "weak" and "strong" betas, as follows:

$$\xi_y \propto \frac{\beta_{yw}}{\sqrt{\varepsilon_{ys} \cdot \beta_{ys}}} \tag{8}$$

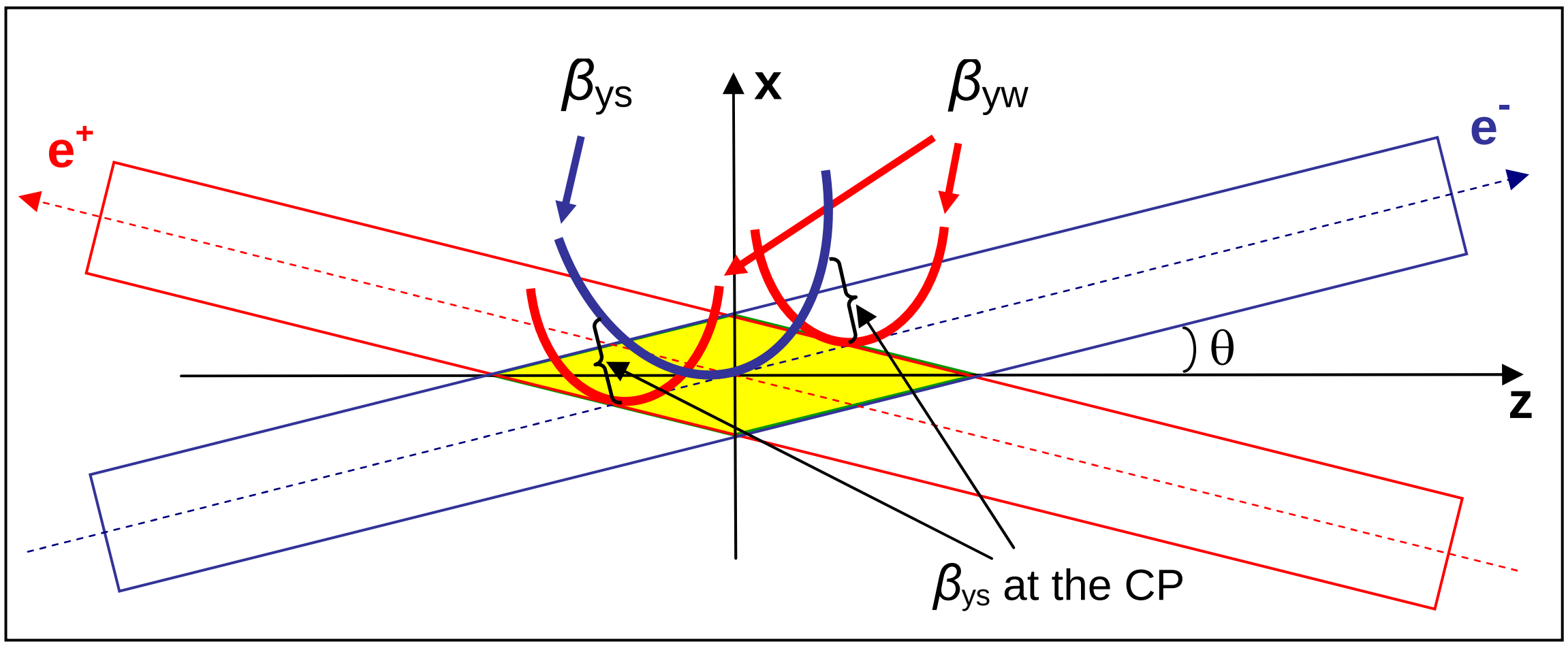

**FIG. 8:** In CW scheme the "weak" $\beta_y$ at the CP's is always in the minimum and does not depend on X-coordinate, but the "strong" $\beta_y$ goes up when CP is shifted from the nominal IP.

Here in the numerator we have "weak" $\beta_y$, and in the denominator – "strong" beam size. Without Crabbed Waist both betas at the CP are actually the same, the difference is negligible when $\theta << 1$. It means that $\xi_y$ scales as $(\beta_{ys})^{1/2}$. In the Crabbed Waist scheme $\beta_{yw} = const$ at the CP, so $\xi_y$ scales as $(\beta_{ys})^{-1/2}$, that is inverse dependence of the one without CW, see Fig. 8. This means that if the waist rotation is smaller than the nominal value, the amplitude modulation should decrease while some phase modulation appears again. From here we can conclude that there is some "optimum" angle of the waist rotation – as a compromise between amplitude and phase modulations, which should depend on the other parameters ($\xi$, $\phi$, etc.). Usually the optimum lies somewhere in the range of 0.6 to 0.8 of the nominal value, see Fig. 9.

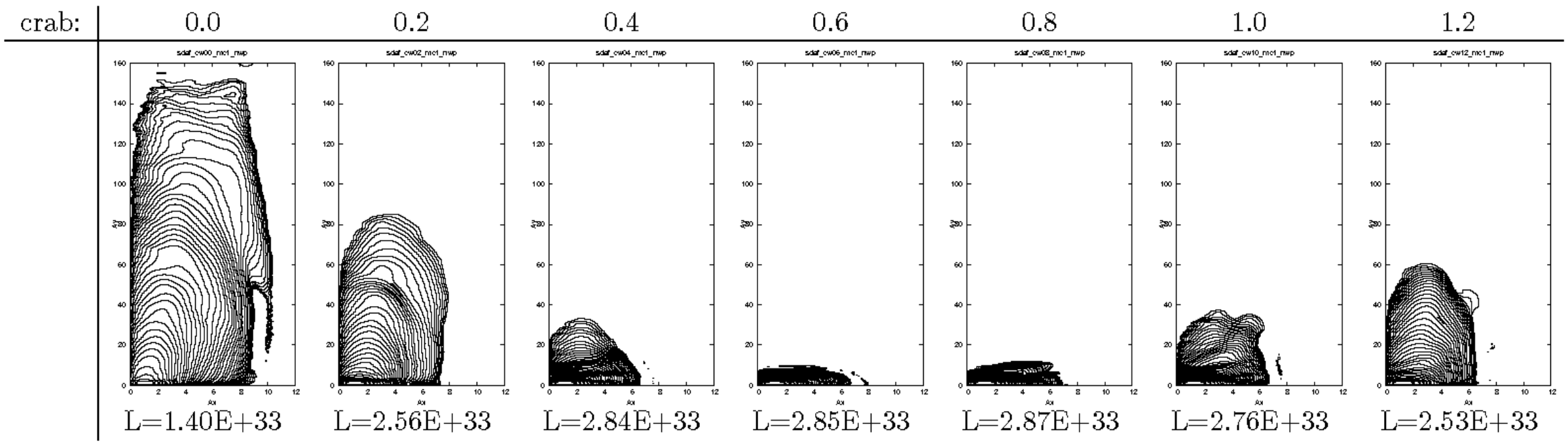

**FIG. 9:** Luminosity and beam tails for SIDDHARTA[3)] vs. waist rotation.

In simulations, however, we can exclude the $\beta_{ys}$ modulation very easily: increase $\beta_{ys}$ by a factor of, say, 100 and decrease the "strong" vertical emittance $\varepsilon_y$ by the same factor. In this case the optimum waist rotation must be shifted to the nominal value 1.0, that is exactly what we obtained in our simulations, see Fig.10. The second row in the table corresponds to 1.5 times higher tune shift ("strong" bunch current).

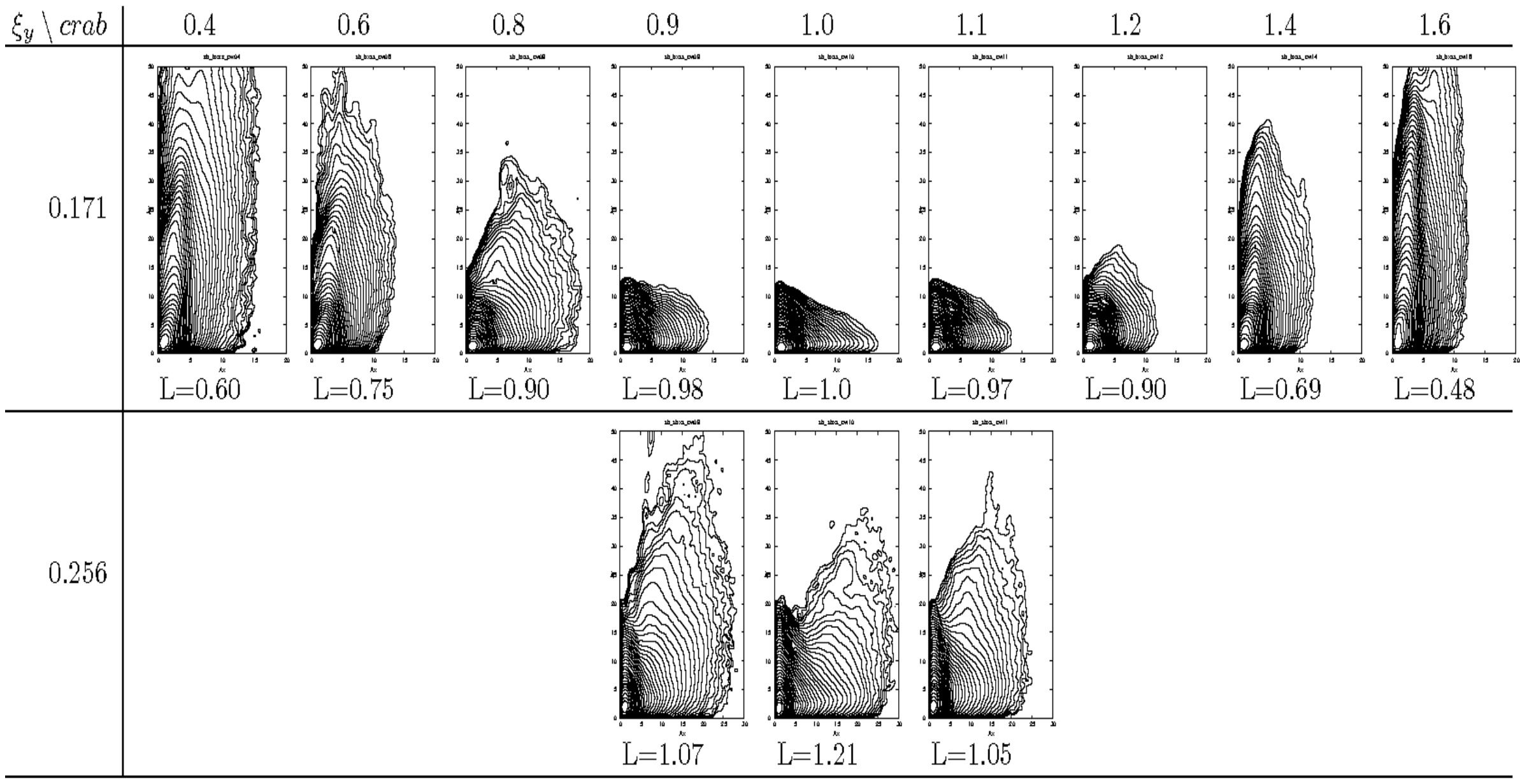

**FIG. 10:** Luminosity in some units and beam tails for SuperB[4)] vs. waist rotation. The "strong" $\beta_y$ is larger and $\varepsilon_y$ is smaller by a factor of 100, so there is no $\beta_y$ modulation.

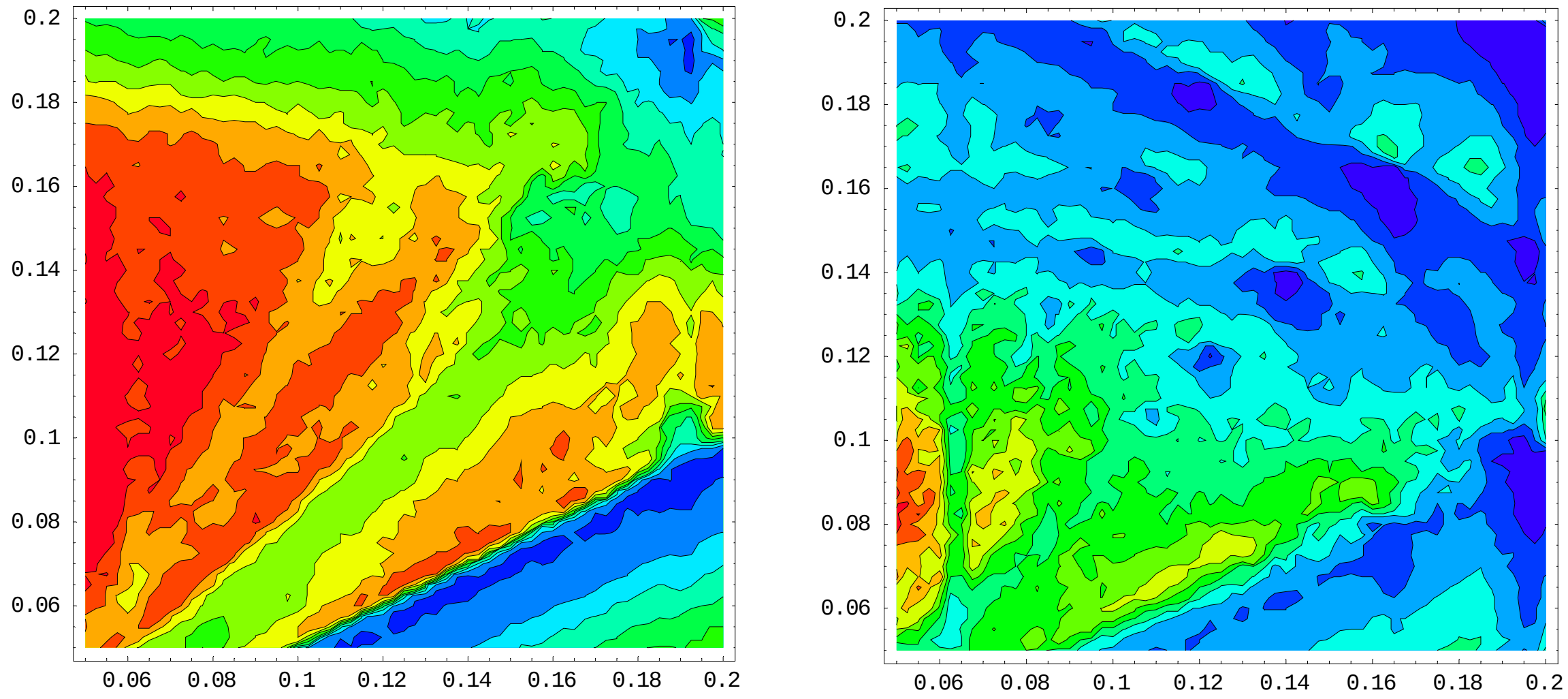

**Crab ON => 0.6/θ**
$L_{max} = 2.97\text{x}10^{33}\ cm^{-2}s^{-1}$
$L_{min} = 2.52\text{x}10^{32}\ cm^{-2}s^{-1}$

**Crab OFF**
$L_{max} = 1.74\text{x}10^{33}\ cm^{-2}s^{-1}$
$L_{min} = 2.78\text{x}10^{31}\ cm^{-2}s^{-1}$

**FIG. 11:** SIDDHARTA luminosity scan[3]. Red colour corresponds to the maximum luminosity, blue – to the minimum.

The effect of the betatron resonances suppression by the CW becomes the most obvious when looking at the luminosity scan vs. betatron tunes, see Fig.11. As one can see, with the Crabbed Waist many X-Y betatron resonances disappear or become much weaker, so the good working area enlarged significantly, plus the maximum luminosity increased by a factor of about 2. One more example is presented on Fig. 12, where the SuperB[4] parameters were used.

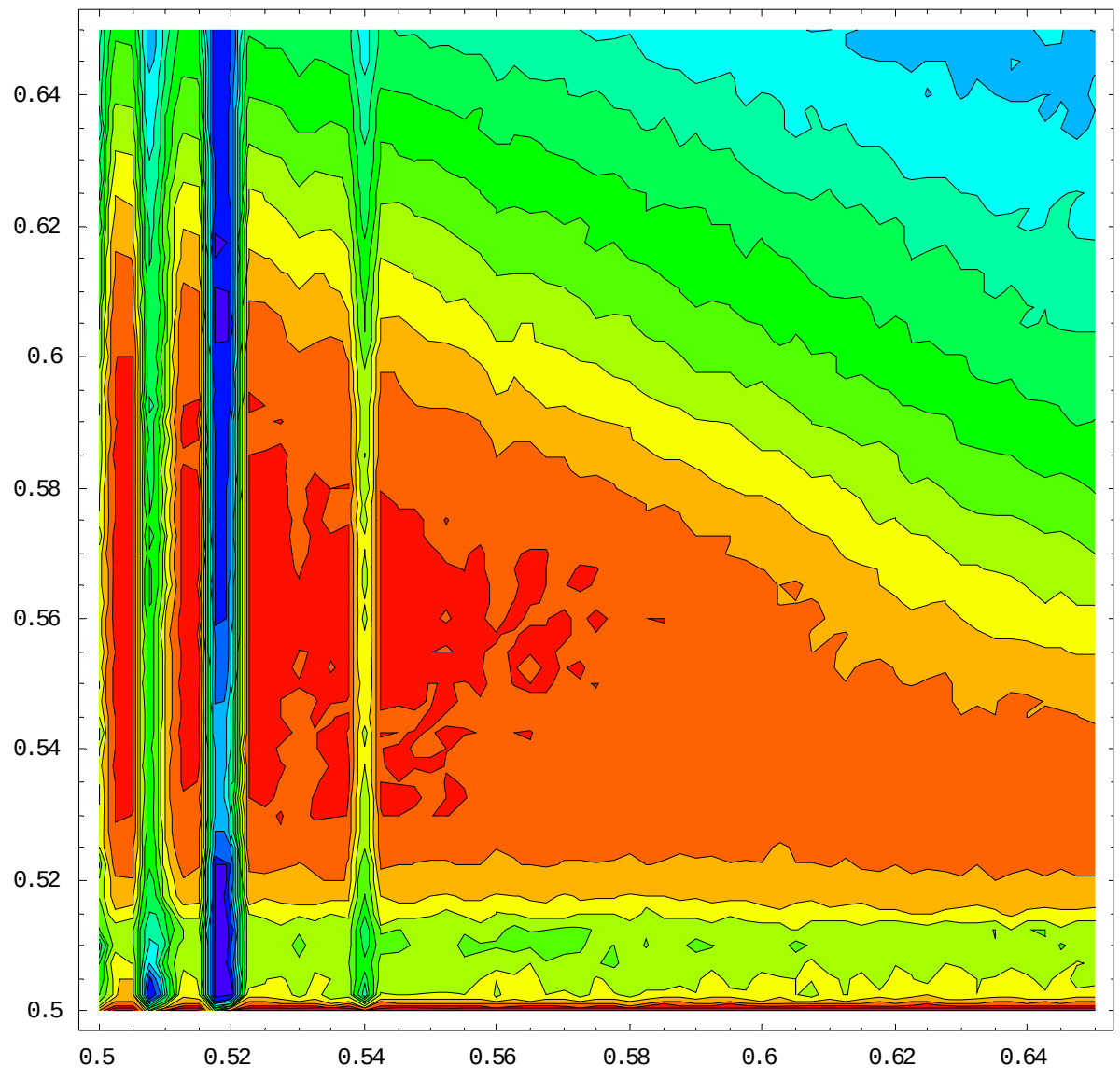

$L_{max} = 1.21\text{x}10^{36}\ cm^{-2}s^{-1}$ $L_{min} = 2.25\text{x}10^{34}\ cm^{-2}s^{-1}$

**FIG. 12:** SuperB luminosity scan[4] with CW, $\nu_s$=0.02.

The only strong resonances remaining are the horizontal synchro-betatron ones (satellites of half-integer resonance). In Fig.11 such resonances (satellites of integer) are not visible, since the scan area there was from 0.05 to 0.20. The vertical synchro-betatron resonances are suppressed, that is in good agreement with[9)]. And we do not see any strong betatron resonances on Fig.12, in contrast with Fig.11, where we can see some. The reason is that the Piwinski's angle for SIDDHARTA is not so big and the CW mechanism is working not in full strength.

## 4 CONCLUSIONS

The main features of the Crabbed Waist scheme of collision can be summarized in three items:

1) Large Piwinski's angle – to decrease the overlapping area.

2) Low $\beta_y$ to fit the overlapping area – this is the main source of the luminosity increase.

3) Crabbed Waist – to suppress the betatron resonances, allows significant $\xi_y$ and luminosity increase.

It is worth to note that the modulations in CW scheme become significantly smaller as compared to head-on collision scheme, thus the beam-beam limit $\xi_y$ increases by a factor of about 2-3!